\documentclass{IOS-Book-Article}
\usepackage{hyperref}
\usepackage{mathptmx}
\usepackage{soul}\setuldepth{article}
\def\hb{\hbox to 11.5 cm{}}

\begin{document}

\pagestyle{headings}
\def\thepage{}
\begin{frontmatter}              

\title{Feeling Guilty Being a c(ai)borg: Navigating the Tensions Between Guilt and Empowerment in AI Use}

\markboth{}{January 2025\hb}

\author[A,B]{\fnms{Konstantin} \snm{Aal}\orcid{0000-0001-7693-7340}%
\thanks{Corresponding Author: Konstantin Aal, konstantin.aal@kgst.de}},
\author[B]{\fnms{Tanja} \snm{Aal}\orcid{0000-0003-4694-4296}}
\author[B]{\fnms{Vasil} \snm{Navumau}\orcid{0000-0003-0087-5248}}
\author[B,C]{\fnms{David} \snm{Unbehaun}\orcid{0000-0001-8551-186X}}
\author[B]{\fnms{Claudia} \snm{Müller}\orcid{0000-0001-8534-546X}}
\author[B]{\fnms{Volker} \snm{Wulf}\orcid{0000-0002-1214-1551}}
\author[B]{\fnms{Sarah} \snm{Rüller}\orcid{0000-0003-2541-2776}}

\runningauthor{K. Aal et al.}
\address[A]{KGSt - Kommunale Gemeinschaftsstelle für Verwaltungsmanagement, Germany}
\address[B]{University of Siegen, Germany}
\address[C]{Clausthal University of Technology, Germany}

\begin{abstract}
This paper explores the emotional, ethical and practical dimensions of integrating Artificial Intelligence (AI) into personal and professional workflows, focusing on the concept of feeling guilty as a 'c(ai)borg' - a human augmented by AI. Inspired by Donna Haraway's Cyborg Manifesto, the study explores how AI challenges traditional notions of creativity, originality and intellectual labour. Using an autoethnographic approach, the authors reflect on their year-long experiences with AI tools, revealing a transition from initial guilt and reluctance to empowerment through skill-building and transparency. Key findings highlight the importance of basic academic skills, advanced AI literacy and honest engagement with AI results. The c(ai)borg vision advocates for a future where AI is openly embraced as a collaborative partner, fostering innovation and equity while addressing issues of access and agency. By reframing guilt as growth, the paper calls for a thoughtful and inclusive approach to AI integration.
\end{abstract}

\end{frontmatter}
\markboth{April 2025\hb}{April 2025\hb}

\section{Introduction}
Artificial Intelligence (AI) is becoming an integral part of our daily lives [1] , reshaping the way we work, communicate and solve problems [2]. From personalised recommendations on streaming platforms to advanced language models assisting with complex tasks or summarizing news and texts [3], AI is no longer confined to science fiction or niche applications - it is ubiquitous [4,5,6]. But despite its ubiquity, AI is not a monolithic entity. It encompasses a diverse range of approaches, technologies and applications, each with unique capabilities, limitations and ethical implications. This diversity requires a nuanced understanding of what AI really is and how it works. By fostering this understanding, we can counter the dominant tech-centric narratives and ensure that AI technologies reflect and respect the pluralistic nature of our global society [7,8]. The ultimate goal is to promote AI systems that are technologically advanced, ethically sound and socially inclusive, marking a significant shift towards a more just and humane technological future [4,8,9].

As we integrate AI into our lives, we are faced with fundamental questions about its role, its impact, and our relationship with it. Perceptions of AI often oscillate between extremes: an all-powerful force capable of revolutionising humanity, or a dystopian tool that will erode our individuality and creativity. Neither perspective fully captures the reality. What is missing is a framework for thinking critically and constructively about AI's place in our lives - one that acknowledges both its potential and its challenges.

This paper introduces the concept of examining AI through the lens of guilt — a powerful emotional and ethical dimension often overlooked in discussions about technology [10]. Guilt, as we propose, is not merely a by-product of using AI, but a lens through which we can better understand our hesitations, responsibilities, and aspirations. For instance, many feel guilty relying on AI to augment creative or intellectual tasks, perceiving this as a compromise of authenticity or effort. Yet, this guilt also reflects deeper anxieties about how we define human ingenuity and value in an age of machine augmentation.

Alongside this lens, we argue that there is an urgent need for a coherent vision of how AI should evolve in both private and professional spheres. Without such a vision, we risk either underutilizing AI’s potential or succumbing to its unchecked proliferation. A thoughtful approach requires us to define the skills, literacies, and frameworks necessary for meaningful and responsible engagement with AI. This vision is particularly relevant in academia and research, where AI challenges traditional notions of intellectual labor and originality.

\subsection{Donna's Cyborg}
Donna Haraway's seminal work, A Cyborg Manifesto (1985), challenged the traditional binaries that have long structured human thought: man versus machine, nature versus culture, physical versus digital. Haraway reimagined the cyborg - a hybrid of organism and machine - as a metaphor for breaking down these rigid distinctions and embracing fluidity and interconnectedness. This perspective was not only technological but deeply political, questioning the boundaries of identity, gender, and knowledge. Haraway's cyborg exists in a liminal space, refusing the constraints of dualism and instead embracing a spectrum of possibilities for being and acting in the world.

In many ways, artificial intelligence embodies Haraway's vision of the cyborg, but with new implications for how humans relate to tools, technology, and themselves. Today, AI is not just a tool we wield, but a collaborator, an integrated part of how we think, create, and solve problems. This hybridisation challenges us to rethink where human creativity ends and machine assistance begins. This evolution has given rise to the concept of the '\textit{c(ai)borg}' - AI as not just an external system, but a unifying element, an augmented extension of human capabilities.

While Haraway's cyborg was a critique of patriarchal, capitalist and militarist frameworks, the \textit{c(ai)borg} reimagines these critiques for the 21st century, focusing on the integration of AI into everyday life. The \textit{c(ai)borg} moves us away from the binary view of man versus machine and towards a collaborative framework in which technology becomes an augmentative partner. This perspective is particularly valuable in academic and creative spaces, where the boundaries between human labour and machine contribution are increasingly blurred [2], such as music production [11], co-writing [12,13], painting together\footnote{\url{https://www.interaliamag.org/interviews/sougwen-chung-human-machine-collaborations/}}, course-design\footnote{\url{https://blogs.lse.ac.uk/highereducation/2024/01/19/can-ai-co-design-the-perfect-course/}}, content creation [14] and more [15,16,17].

Haraway’s work also encourages a move away from guilt-laden narratives around technological dependence. In her manifesto, dependence on machines is not a weakness but a redefinition of identity and capability. Similarly, the \textit{c(ai)borg} suggests that reliance on AI need not be framed as a loss of authenticity or human ingenuity. Instead, it can be seen as an expansion of what it means to be human — a step toward a more interconnected and augmented future.

In this vision, AI functions as a connective tissue in the digital age, enabling collaboration, creativity, and problem-solving in ways that were previously impossible. Like Haraway’s cyborg, the \textit{c(ai)borg} rejects the notion that purity or independence defines humanity. Instead, it embraces hybridity as a strength and a necessity in navigating the complexities of a rapidly changing world.
\subsection{Contribution}
Following this introduction, we provide a current overview of the role of AI in everyday life, focusing on its increasing ubiquity and the ethical considerations this raises. We then explore the concept of guilt and how it manifests in our relationship with AI. Next, we outline a methodological approach that uses auto-ethnography to reflect on personal experiences with AI. In the findings section, we discuss the dual dimensions of how AI is used and the emotions it evokes. Finally, the discussion considers the wider implications of these findings, in particular the skills needed to navigate the integration of AI into our lives, before concluding with a vision for a future where AI serves as an augmentative and transparent tool.

At the heart of this paper is the belief that AI is not just a tool, but a transformative force that requires us to rethink how we work, create and live. By confronting the complexities of our relationship with AI - including the guilt it can provoke - we hope to contribute to a broader understanding of how AI can enrich rather than diminish the human experience. The way forward is not to resist AI, but to learn to engage with it critically, openly and ethically. This paper seeks to begin the process of rethinking current barriers and prejudices in the usage and to start a conversation about the need to reflect on the future of AI involvement during various processes (e.g. academic writing).

\section{State of the Art}
The next sub-chapters explore key aspects of AI’s evolving role. The first, AI literacy, focuses on understanding AI’s capabilities and limitations for responsible use. The second, Guilt, examines the emotional impact of relying on AI in daily and professional life. Together, they provide a framework for integrating AI effectively.
\subsection{AI Literacy}
AI literacy is increasingly recognised as a critical competency for individuals navigating a world shaped by artificial intelligence [9]. It goes beyond the ability to use AI tools to include a deeper understanding of how AI works, its societal implications, and the ethical considerations surrounding its use. Researchers such as Long et al. [18], Chiu et al. [19], and others [20,21] have emphasised AI literacy as a multifaceted set of skills and knowledge that enables individuals to critically evaluate, communicate, and collaborate effectively with AI systems.

AI literacy requires understanding core AI concepts, including its types (narrow, general, superintelligent AI) and machine learning approaches (supervised, unsupervised, reinforcement learning) to grasp AI’s capabilities and limitations [21]. Data literacy is also crucial, enabling individuals to interpret and critically assess data used in AI training [21]. Additionally, algorithmic literacy involves evaluating algorithms’ reliability, recognizing biases, and determining when to trust or question AI-driven decisions [22].

AI literacy also includes awareness of the broader societal implications of AI. This includes recognising both the positive and negative impacts of AI technologies, such as potential biases in AI systems, issues of transparency, and the evolving role of humans in AI development [21,22]. Understanding these societal dynamics is critical to promoting responsible engagement with AI. Ethical considerations are central to AI literacy. Individuals need to be aware of issues such as fairness, privacy and accountability in the use of AI. Ethical literacy ensures that users can critically assess how AI aligns with values such as equity and justice, minimises harm, and promotes responsible use [8,19,21,22]. Another emerging component of AI literacy is prompt engineering - the ability to design effective prompts to guide AI systems towards desired outcomes. As AI tools such as language models become more integrated into professional and personal workflows, prompt engineering is increasingly recognised as a critical skill for maximising their utility [6,21,23]. AI literacy is not a static skill, but an evolving process. As AI technologies continue to advance rapidly, continuous education and adaptation is required to stay informed about new developments and their implications [9,19,22,24].
\subsection{Through the Lens of Guilt}
Research on guilt has a long tradition in participatory design [10,24,25,26], where it is often framed as a response to unfulfilled obligations or injustices towards design partners. Far from being an uncomfortable feeling, guilt can serve as a powerful lens for examining the ethical complexities of participatory design research, particularly when working with marginalised or vulnerable communities. It provides researchers with a framework for critical self-reflection, encouraging them to confront the ethical dimensions of their work and to strive for more equitable and impactful practices [10].

Guilt from unmet expectations underscores the need to align research with community needs, promoting co-creation over external agendas [25]. Guilt over exploitation highlights power dynamics, encouraging equitable, mutually beneficial practices [26]. Concerns about professional recognition reveal tensions between personal success and community well-being, shifting focus to lasting impact. Similarly, guilt over limited outcomes fosters a deeper understanding of participatory research, emphasizing systemic change and long-term engagement over short-term results [27].

Hwang et al. explore AI’s impact on higher education, emphasizing AI literacy, rapid engineering, and critical thinking as essential skills. Using a Swiss university case study, they offer strategies for responsible AI integration in teaching and learning [28]. Similarly, Draxler et al. examine the “AI Ghostwriter Effect,” where users of AI text tools neither claim authorship nor credit AI but are more likely to acknowledge human ghostwriters [29]. Concerns about authenticity and ownership arise [12]. Khosrowi et al. address AI-generated content attribution, proposing a “collective-centered creation” (CCC) framework that distributes credit among users, AI systems, developers, and data providers based on their contributions [30].

Acknowledging and addressing guilt is crucial to fostering ethical growth in the use of AI systems. Rather than seeing guilt as merely a negative or paralysing emotion, it can be reframed as a constructive force or lens that encourages critical reflection on the use of AI and the responsibilities it entails. This perspective aligns with the paper's emphasis on transparency, accountability and responsiveness, and highlights how engaging with guilt can lead to more thoughtful and equitable interactions with AI tools. Confronting the discomfort associated with AI use allows users to critically assess the ethical implications of their reliance on these technologies, challenge assumptions about their capabilities, and ensure that AI applications are aligned with broader goals and values.

This approach also underscores the importance of cultivating basic and advanced skills, such as critical thinking and AI literacy, to evaluate AI outcomes responsibly and effectively. By moving from guilt to empowerment through transparency and skill-building, users can reframe AI as a collaborative partner rather than a source of ethical discomfort. In this way, guilt acts as a catalyst for ethical engagement, in line with the manuscript's broader call for a reflective and ethically conscious approach to integrating AI into everyday and professional life.

\section{Methodological Approach}
Autoethnography, as described by Adams et al. and Ellis et al., is a qualitative research method that combines autobiography and ethnography to examine cultural phenomena through personal experience. Using their lived experiences as primary data, autoethnographers critically reflect on the interplay between their individual narratives and broader social or cultural practices [31,32].

This approach involves introspection and self-reflection, drawing on diverse materials such as memories, personal diaries and dialogues to generate nuanced insights into the complexities of human-AI interaction [33]. By embracing vulnerability and fostering reciprocity with audiences, autoethnography disrupts traditional norms of research practice, offering insider knowledge while working towards greater inclusivity and positive change [31]. However, it also raises ethical concerns about the representation of personal and shared experiences, requiring careful consideration of how narratives are constructed and shared [34,35].

In this context, autoethnography provides a valuable framework for examining the emotional, ethical and practical dimensions of AI use, offering insights into the complex interplay between personal agency, cultural expectations and technological capabilities. This study uses an autoethnographic approach to explore the authors' personal experiences and reflections on the use of AI systems and tools. By grounding this exploration in the authors' emotions and observations, the study aims to explore the complex relationship between users and AI, focusing in particular on feelings of guilt, excitement and ethical considerations.

One of the authors has been an extensive user of AI since 2023, constantly experimenting with the latest tools (including ChatGPT\footnote{\url{https://chatgpt.com/auth/login}}, Quivr\footnote{\url{http://www.quivr.com/}}, Midjourney\footnote{\url{https://www.midjourney.com}}, DALL-E\footnote{\url{https://openai.com/index/dall-e-3/}}, NotebookLM\footnote{\url{https://notebooklm.google.com/}}, Claude\footnote{\url{https://claude.ai/}}, Perplexity\footnote{\url{https://www.perplexity.ai/}}). Initially, his engagement with AI was driven by curiosity and a sense of excitement about its potential. AI was used primarily for recreational and exploratory purposes, with little fear that it would replace human jobs. Instead, AI was seen as a promising tool for enhancing human creativity and productivity. However, this period was also marked by a conflicting sense of guilt - an emotional response to the use of AI, even when it seemed harmless or playful. Over time, as the author incorporated AI into more practical and professional contexts, the initial guilt began to evolve, giving way to a more nuanced understanding of AI's role in human work.

In contrast, another author was and is much more cautious about AI, using it more often since 2024. Although she recognises the potential benefits and has explored systems such as ChatGPT, DALL-E, and NotebookLM, she feels deeply uncomfortable using AI for everyday or professional tasks. This reluctance stems from ethical concerns, particularly in a professional context, a strong attachment to traditional notions of creativity and authenticity, and a lingering sense of guilt when AI is used for tasks in systems whose societal frameworks do not appear to offer the possibility of transparency, particularly due to the prevalence of ego-centred mechanisms and the coupling of performance, acknowledgement and identity. 

Yet another author was hesitant to use AI applications due to uncertainties surrounding the usage of personal data, rarely using it 2023 and more often since the middle of 2024. However, positive experiences of colleagues convinced him to explore these tools (such as Chat GPT, NotebookLM, Revoldiv) in search of a more balanced approach to the writing process. Engaged in a variety of writing assignments, he was also excited by the possibility of using AI to document his brainstorming sessions and translate them into coherent texts. At the same time, he felt uneasy by the standardized responses generated by AI, overshadowing his nuanced but imperfect writing style and guilty for delegating tasks he previously handled himself to AI. His experience now is marked by the necessity of navigating these two extremes – balancing authencity with efficiency, a crucial demand of his academic work.          

This difference in perspective between the authors sparked numerous conversations about the possibilities and emotional impact of AI. These discussions became a cornerstone of this manuscript, highlighting the different ways in which individuals engage with and interpret AI technologies.

The idea for this paper emerged from these long, reflective conversations between the authors, focusing on the emotional and ethical dimensions of AI use. To deepen the analysis, the authors drew on their own observations, capturing the shifts in their attitudes and behaviours as they navigated AI tools in both personal and professional contexts. These experiences were further informed by conversations with colleagues, a review of relevant literature, and direct experimentation with various AI applications.
\section{Findings}
This section presents three examples of how AI tools are used in practice by three authors of this paper, highlighting the diverse ways AI is integrated into both professional and personal workflows. These examples delve into the emotional and practical dimensions of AI usage, offering insights into how these tools augment creativity, streamline tasks, and provoke reflection on their broader implications, but also show how inappropriate and unethical the same tools can be used for. These are not personas, but real use cases and practices from an academic context. Other contexts such as industrial roles can differ from the experiences described here.
\subsection{Embracing the Use of AI}
“AI has significantly transformed both my professional and personal life. One of its most valuable roles is as an ideation and conversation partner. I often start with a rough idea and use AI to refine, expand, or explore alternative perspectives. This interaction sharpens my thinking, encourages reflection, and consistently helps me expand on ideas.

In my personal life, AI supports me in drafting blog posts. It’s not just about writing but also brainstorming and refining. I sketch out the structure of an article and ask AI to suggest new aspects or alternative formats, acting as a creative partner to refine and expand my work.

Professionally, AI has revolutionized my research process. Tools like Notebook LM help me quickly grasp the core arguments of academic papers or summarize educational videos, allowing me to prioritize materials worth deeper attention. Similarly, AI enables me to extract key takeaways from my backlog of saved YouTube videos, streamlining the incorporation of insights into my ideas.

AI has also enhanced my note-taking and brainstorming. Tools like Otter.ai let me record voice notes on the go, capturing fragmented thoughts that AI later helps me refine into coherent drafts. This workflow eliminates writer’s block and aligns with my non-linear thinking, turning scattered ideas into productive outputs.

Integrating AI into my workflow has been transformative, expanding my thinking, articulating ideas, and optimizing processes for creativity and efficiency. However, this comes with a sense of guilt. AI accelerates tasks that many still handle manually, raising ethical questions about reliance on these tools and their impact on traditional practices.

Despite this guilt, the benefits are undeniable. AI has reshaped how I approach tasks, making creativity and productivity more accessible and fluid than ever before.”
\subsection{Shame and Guilt in the Use of AI}
"After preparing a formal response to three conference presentations using only brief summaries, I found myself deeply reflecting on AI, shame, and guilt. The task, set within an academic environment demanding theoretical depth and precise language, was daunting given the limited data and lack of clear expectations. My attempts to find guidance online were unsuccessful and consumed precious time. Turning to ChatGPT, I provided the summaries and received a polished draft that exceeded expectations, requiring only minor edits. However, this left me conflicted: should I use the AI-generated response or start again? Despite rationalizing my choice to use it, I felt uneasy, as the task I was paid to perform had been largely completed by AI.

After delivering the response, which received overwhelming praise, I felt guilt rather than pride. The praise felt undeserved, and I struggled with the lack of transparency about AI’s role. Reflecting on this, I realized the discomfort stemmed from societal expectations around intellectual work, authenticity, and transparency. This experience reminded me of earlier debates about tools like DeepL, now widely accepted for translations but initially controversial. Generative AI like ChatGPT faces greater scrutiny as it intersects with core creative and intellectual processes tied to identity and professional expectations.

The lack of clear frameworks for using AI in professional settings exacerbates these conflicts. In my case, openly acknowledging AI’s role risked reputational damage. Invisible uses of AI, like summarizing or idea generation, are easier to accept, while visible outputs, like drafting formal responses, provoke more scrutiny.

This experience raises broader questions: When is it acceptable to use AI to improve efficiency? How can we establish legitimacy and transparency for AI in public and professional contexts? Negotiating AI’s role is an ethical, societal, and personal process. My reflection is part of a larger dialogue on how we navigate this transformative technological shift, balancing AI’s potential to enhance creativity with the need to preserve authenticity and transparency.”
\subsection{Balancing Efficiency and Authenticity in AI Use}
“When I think about using AI tools, it’s not just about convenience—it’s tied to my identity, responsibilities, and the integrity of my work. Take translating or transcribing interviews, for example. AI tools like Revoldiv and DeepL have transformed this process, cutting hours of work into just one. This efficiency allows for more time on deeper analysis, but it comes with a sense of loss.

Whenever I use AI to translate or polish texts, I feel guilt. As someone with a hybrid identity rooted in Eastern Europe, I feel responsible for upholding the integrity of my native language, especially since Belarusian is considered vulnerable. Handing this task to AI feels like neglecting a part of myself. I wonder if I should ensure my thoughts are expressed with the cultural nuance only I can provide.

Yet, working in English-dominated spaces as a non-native speaker, I rely on AI to make my writing clear and polished for high-level journals. While it helps meet these standards, it often strips away nuances tied to my local context, making my work feel less personal and layered. This trade-off has made me more cautious about using AI for creative tasks. AI operates within existing patterns and lacks the originality rooted in unique perspectives, shaping my work in ways that don’t always align with my intentions.

Now, I use AI selectively—for polishing drafts, translating ideas, or composing quick emails. While it’s invaluable for straightforward tasks, I still wrestle with guilt. Am I compromising my skills or authenticity? Can I balance efficiency with preserving my voice? These questions push me to reflect on my writing process, address weaknesses, and consider how AI fits into this journey. Yet, AI can never replicate the brainstorming sessions with colleagues, where the exchange of ideas fuels both creativity and collaboration—the heart of academia.”
\section{Discussion}
The integration of AI into creative, academic and professional spheres has sparked significant debate, particularly around the emotional and ethical implications of its use. Concerns about authenticity, skill erosion, and societal perceptions highlight the complexities of adopting AI as a collaborative partner. This discussion explores these challenges, focusing on fears of 'cheating', judgement by others, and the potential loss of personal skills.

As outlined in previous sections, these fears are deeply intertwined with broader societal narratives and the evolving role of AI in reshaping human endeavour. By examining these fears through the lens of guilt and the frameworks of AI literacy and ethical engagement, we can better understand the tensions that arise in balancing augmentation with personal contribution. These reflections aim to reframe the conversation around AI, moving from fear to empowerment and advocating for the thoughtful, transparent and intentional use of AI technologies. Through this lens, we can envision a future where AI enhances creativity and productivity while preserving the authenticity and uniqueness of human endeavour.
\subsection{Fear of Cheating}
For many, using AI for tasks feels like cheating, especially in writing, which is traditionally seen as a deeply creative and intellectual process requiring thought and originality. AI's ability to generate content instantly can feel like bypassing essential human contributions, reinforced by societal narratives that equate authenticity with individual effort.

This guilt reflects wider concerns about fairness, authenticity and responsibility in the use of AI [10,21,24]. Studies such as Draxler et al.'s "AI ghostwriter effect" show that users are reluctant to acknowledge AI contributions for fear of diminished authenticity and credit [29]. These issues highlight the tensions between human effort and AI augmentation.

AI literacy, as discussed by Long et al. [18], Stolpe and Hallström [22], and others [8,21], provides tools to address these concerns. It reframes AI as a collaborative partner, useful for brainstorming, refining ideas and organising thoughts - tasks that complement rather than replace human creativity. Ethical practices that emphasise transparency and accountability help alleviate discomfort and empower users.

Expanding AI literacy to include prompt engineering and critical evaluation, as noted by Hwang et al. [6], ensures that users intentionally engage with AI and adjust outputs to maintain originality and intellectual rigour. Cultivating these competencies allows users to move from guilt to empowerment, seeing AI as a partner that enhances creativity and intellectual work while addressing the ethical challenges of an AI-enhanced world.
\subsection{Judgment by Others}
There is a widespread fear of being judged or punished for using AI, as many fear that admitting reliance on AI could lead to negative perceptions of their abilities. Students, for example, fear being accused of academic dishonesty, while professionals such as writers or marketers may be reluctant to disclose their involvement with AI for fear it will devalue their work or creativity. In the workplace, employees often worry that using AI will be seen as lazy or incompetent, fostering a culture of secrecy and limiting AI's potential [10,12].

This reluctance stems from societal narratives that equate individual effort with authenticity and intellectual worth. Addressing these fears requires fostering AI literacy and changing cultural perceptions, as emphasised by Long et al. [18] and Stolpe and Hallström [22]. AI literacy involves not only understanding AI, but also engaging critically and transparently with its applications, recognising it as a tool for collaboration and augmentation. Framing the use of AI as a skill rather than a shortcut reflects adaptability and competence, as noted by Hwang et al. [6] and Ryan [8].

Systemic changes in education and workplaces are essential. Policies that promote AI literacy and ethical use can reduce stigma and foster a culture where AI is seen as a collaborative partner, rather than a source of fear. This shift is critical to unlocking the potential of AI in personal, academic and professional contexts, making it a tool for empowerment [21,22].
\subsection{Losing my own Skills}
Creativity is deeply personal, reflecting an individual’s unique voice and perspective. However, as AI increasingly handles creative tasks, concerns about originality and the erosion of human creative skills arise. AI, designed to mimic patterns from vast data, reshapes intellectual and creative work, raising fears that over-reliance on it could weaken personal skills and authenticity [1,2,8,21].

AI literacy, as emphasized by Long et al. [18] and Stolpe and Hallström [22], is vital for navigating this tension. Understanding AI’s capabilities and limitations allows users to consciously integrate it as a collaborative partner, ensuring their voice remains central. AI can enhance creativity without replacing personal effort, enabling creators to focus on conceptual work while leveraging AI for efficiency.

Broader societal concerns about authenticity and intellectual labor underscore the need for transparency and ethical use [12,29]. Expanding AI literacy to include skills like prompt engineering and critical engagement empowers individuals to use AI effectively while preserving their unique contributions [6,21]. By fostering ethical AI practices, users can embrace these tools as a means of augmentation and empowerment, ensuring that human creativity remains at the forefront.
\section{Skills and Implicatons for the Future}

\subsection{AI Literacy and Prompt Literacy}
AI and prompt literacy are essential skills for effective engagement with artificial intelligence systems, and include both understanding how AI works and knowing how to interact with it to achieve desired outcomes. AI literacy involves understanding fundamental concepts such as the algorithms that drive AI, the differences between systems such as rule-based models and machine learning, and the critical role that data plays in shaping AI outcomes. This understanding will enable users to select the right tools for tasks and to critically evaluate results, especially given that biases in training data can lead to flawed outputs [8,21,22]. Awareness of ethical considerations, such as fairness and privacy, is also an important aspect to ensure responsible use in line with societal values [6].

Prompt literacy focuses on creating effective queries to communicate with AI tools. Clear and specific prompts are key to eliciting accurate responses, and users need to experiment with phrasing to refine results. This iterative approach maximises the potential of AI while maintaining a critical stance towards its limitations [6,8,21]. Together, these literacies enable individuals to move beyond passive use and use AI as a collaborative tool in ways that are ethical, innovative and purposeful. Developing these literacies is critical to harnessing the power of AI while navigating its complexities, whether in personal or professional settings. By fostering both AI and prompt literacy, users can ensure that AI serves as a tool that enhances rather than replaces human judgement and creativity.
\subsection{Critical Engagement, Honesty and Transparency}
AI is a powerful tool for augmenting human efforts but must never replace human judgment, creativity, or core contributions. Its greatest value lies in managing repetitive or time-consuming tasks, allowing individuals to focus on strategic and creative elements [8,21]. However, critical engagement with AI requires an understanding of its limitations. AI-generated outputs, while often convincing, may include inaccuracies, biases, or fabricated information, necessitating scrutiny, fact-checking, and alignment with intent and [6,12,21]. Blind trust in AI risks perpetuating errors and undermining authenticity and ethics.

Preserving human creativity and originality is essential. While AI can streamline workflows and generate ideas, it must not overshadow the human touch that gives work meaning. By delegating mechanical tasks to AI, individuals can focus on innovation and personal contributions, ensuring that human ingenuity remains central [21]. Honesty and transparency are crucial when using AI. Acknowledging when and how AI has been used fosters trust, accountability, and ethical engagement. Embracing AI reflects adaptability and enhances collaboration between humans and technology. Transparency normalizes AI’s role as a tool for efficiency and creativity, rather than perpetuating the illusion of sole human effort [6,8,21,22].

To foster transparency in the use of AI, our social and academic systems must evolve to create safe and supportive spaces where individuals can openly discuss their reliance on these tools. Normalising conversations around AI use, and integrating clear guidelines for ethical engagement, can help reduce the stigma and fear associated with 'coming out' about AI dependency. These spaces should encourage honesty, celebrate adaptability, and emphasise the value of collaboration between human creativity and machine assistance.
\subsection{Augmentation and Collaboration}
AI is best understood as a tool that enhances human creativity and productivity, similar to a spell checker or grammar assistant. Rather than replacing human ingenuity, it serves as a collaborator that streamlines processes, handles repetitive tasks and provides fresh perspectives. This partnership allows users to focus their energy on higher-level creative and strategic decisions, leveraging AI to increase efficiency and output [8,21].

AI can help refine ideas, suggest improvements or generate initial drafts, but direction, intent and originality must remain the responsibility of the user. This approach ensures that the final product retains a personal, human touch in line with the creator's unique vision [6,22]. By collaborating with AI, users preserve the authenticity and intent behind their work while maximising efficiency.

AI also encourages experimentation and innovation. As a brainstorming partner, it can suggest new angles or unexpected solutions that users may not have considered. This dynamic interplay between human and machine creativity can lead to richer, more innovative results than either could achieve independently [12,18].

By framing AI as a collaborator, users can harness its strengths to work faster and more effectively, without compromising their creative direction or sense of ownership. This mindset ensures that AI becomes an empowering partner in the creative process, enabling users to achieve results that are efficient and distinctly human [8,21].
\subsection{Curiosity in AI Use}
Curiosity is often the engine driving initial engagement with AI, as also seen in the three testimonials. One is intrigued by what these new tools can do, how they respond, and how they might help users and be more effective or faster. This initial curiosity leads to exploration, experimentation, and learning of the capabilities but also limitations of AI. The authors in the provided paper experienced this initial phase, driven by excitement about AI's potential.

If AI systems are designed solely for speed and providing the most probable answer, which they are trained for, they risk amplifying our cognitive shortcuts and limiting exposure to diversity. To counteract this, AI design should intentionally foster deeper, more effortful curiosity by proactively surfacing alternatives and nuance [5,37,38]; for instance, created answers or text could offer contrasting viewpoints ("Here's a common perspective... however, an alternative view argues..."), highlight complexity ("This is a simplified overview..."), or indicate source diversity ("Primarily based on Type A sources... perspectives from Type B might differ..."). Introducing "productive friction" through clarifying questions or requiring minor refinement can also encourage more thought, rather than passive acceptance [39]. Furthermore, integrating explicit "Exploration" features like "Challenge this assumption" or "Find Dissenting Opinions", offering varied output styles such as Socratic dialogues or debates, and consistently framing AI output as a starting point rather than a definitive endpoint can nudge and support users towards deeper engagement. Crucially, transparency about the AI's limitations and potential biases is essential to prompt users to seek broader perspectives and exercise critical thinking.
\section{Conclusion, Next Steps and Vision}
The integration of AI into everyday and professional life presents both opportunities and challenges. While it enhances creativity, streamlines workflows and provides innovative solutions, it also raises concerns about ethics, transparency and accountability. To engage effectively with AI, basic skills such as literacy and critical thinking are essential to evaluate AI outputs and ensure they are aligned with user goals.

Advanced skills, including AI literacy, prompt literacy, and critical engagement, are critical for ethical and informed use. These skills enable users to treat AI as a partner rather than a crutch and to navigate its complexities with confidence and integrity. Moving from guilt to empowerment requires transparency about the role of AI and a collaborative mindset that transforms AI from a challenge into an opportunity for growth and innovation.

Next steps would include moving beyond specific experiences to see if these feelings, transitions, and use cases are shared by a wider, more diverse population across different disciplines, professions, cultural backgrounds, age groups, and levels of AI expertise. In addition, it would be interesting to more rigorously test the proposed relationships, such as the link between increased AI literacy/transparency and reduced guilt/increased empowerment. Based on these considerations, practical strategies or guidelines for individuals and institutions could be developed and possibly tested, building on the call for transparency and literacy.

To ecnourage the transparency approach, we suggest adding an "AI statement" at the end of the paper, similar to the acknowledgements. Here the authors can explain what kind of tools they have used and to what extent.  At the end of this paper, we made a statement explaining the different tools we used and what they were used for.

\subsection{The c(ai)borg}
Inspired by Donna Haraway's Cyborg Manifesto, the vision for \textit{c(ai)borg} emphasises a future in which AI becomes a tool for empowerment, equality and collaboration, rather than a source of exploitation or injustice. Haraway's work highlighted the political implications of technological integration and critiqued the inequalities embedded in the systems of her time. Today, these critiques resonate as we grapple with the ethical challenges posed by AI: Who gets to design AI? Who benefits from its capabilities? How do we ensure that AI serves to augment humanity rather than exacerbate power imbalances?

The \textit{c(ai)borg} envisions a future where these questions are addressed through a thoughtful and inclusive approach to the technology. This future is one in which AI is openly acknowledged as a collaborative partner in our lives, enhancing our capabilities in both personal and professional contexts. Central to this vision is the idea of shared dialogue and exchange. The \textit{c(ai)borg} envisions a world where conversations about AI use are commonplace - where individuals openly discuss how they use AI to support and enhance their work. This openness not only normalises AI as a tool for augmentation, but also encourages collective learning and adaptation, ensuring that AI is used responsibly and effectively.

\newpage
\textbf{AI Statement}
\begin{center}
    During the preparation and writing of this paper, several AI tools were used to enhance the process and streamline certain tasks:
\begin{itemize}
    \item Otter.ai was used to transcribe initial thoughts and the basic structure of the paper. This tool allowed for a seamless transfer of verbal ideas into written form and served as a starting point for further refinement and development.
    \item ChatGPT was used to improve the readability of the findings statements. By rephrasing and organising initial drafts, ChatGPT helped to make the arguments more coherent and accessible, while retaining the original intent.
    \item NotebookLM was used to gain a better overview of the existing literature. This tool provided summaries and insights into relevant sources, helping to contextualise the paper's arguments within the wider academic discourse.
    \item DeepL was used to translate the German transcripts of the findings to English.
\end{itemize}
These AI tools were used as collaborative partners, enhancing creativity and efficiency while maintaining full human oversight to ensure intellectual rigour and authenticity. Their contributions reflect the evolving role of AI in academic research and writing.
\end{center}

\end{document}